\documentclass[aps,prb,twocolumn,showpacs,floatfix]{revtex4}

\usepackage{graphicx}% Include figure files
\usepackage{dcolumn}% Align table columns on decimal point
\usepackage[english]{babel}
\usepackage{amsfonts,amsmath,amssymb,bm}
\begin{document}

\title{Is magnetoresistance in excess of 1,000 \% possible in Ni point contacts?}
\author{A. R. Rocha, T. Archer and S. Sanvito}
\email{sanvitos@tcd.ie}
\affiliation{School of Physics, Trinity College, Dublin 2, IRELAND}
\date{\today}

\begin{abstract}
Electronic transport in nickel magnetic point contacts is investigated with a combination of density functional theory and the non-equilibrium Green functions method. In particular we address the possibility of huge ballistic magnetoresistance in impurity-free point contacts and the effects of oxygen impurities. On-site corrections over the local spin density approximation (LSDA) for the exchange and correlation potential, namely the LDA+$U$ method, are applied in order to account for low-coordination and strong correlations.
We show that impurity-free point contacts present magnetoresistance never in excess of 50\%. This value can raise up to about 450~\% in the case of oxygen contamination. These results suggest that magnetoresistance in excess of 1,000~\% can not have solely electronic origin.
\end{abstract}


\maketitle

\section{Introduction}\label{intro}

Since the discovery of the giant magnetoresistance (GMR) in magnetic multilayers \cite{baibich_gmr,binasch_gmr} there has been an increasing interest in the electronic transport properties of magnetic materials. Besides establishing 
the ground for the manufacture of the present generation of magnetic data storage devices, GMR has also provided a new paradigm whereby the electron spin as well as its charge can be used for electronic applications.\cite{spintronics}

The GMR effect is the overall change in the electrical resistance of a magnetic device when an external magnetic field is applied and it is associated to a change in the internal magnetic structure of the device. For instance, when the magnetic moments of the two magnetic layers forming a spin valve are 
aligned parallel to each other from their zero-field antiparallel alignment, the resistance across the junction drops. This is the principle behind the present generation of read-heads for hard-drives. However, in order to reach storage densities of the order of Terabit/in$^2$ a substantial down-scaling of the read/write devices is needed. One possible avenue to this goal is offered by magnetic point contacts (MPCs), with typical cross sections approaching the atomic scale. In these the electron coherence length is greater than the size of the point contact and the electronic transport is ballistic.

The experimental landscape for magnetic point contact transport is rather rich and controversial. Viret {\it et al.} have performed mechanically controlled break junction experiments in cryogenic vacuum and shown that the ballistic magnetoresistance (BMR) in a one-atom-thick MPC can reach up to 40~\% , \cite{viret} in good agreement with theory.\cite{mertig1}  Other experiments 
using slightly larger constrictions and performed in air showed large ballistic magnetoresistance (LBMR) ratios 
reaching up to 300~\% .\cite{garcia1999,versluijs,oscar1,oscar3} Finally claims of a huge ballistic magnetoresistance 
(HBMR) effect have been recently made with values up to a few thousand percent.\cite{garcia2001,zchopra}

To date there is a long-standing debate around the origin of the LBMR and HBMR in MPCs, which is not
completely resolved. On the one hand, it has been argued that magnetic field-induced mechanical effects 
can produce large magnetoresistance (MR). In fact, either magnetostriction, \cite{viret3,viret06} dipole-dipole 
interactions\cite{chung_dipole} and magnetically induced stress relief\cite{ansermet} may cause the compression 
of the nano-contact once a magnetic field is applied. This enlarges the cross section of the MPC and consequently 
the resistance of the junction decreases. On the other hand, there is also evidence that mechanical effects alone 
are not able to account for whole MR. Specifically Garc\'{\i}a {\it et. al.} have shown that the behaviour of MPCs 
does not comply with mechanical changes, in particular with magnetostriction \cite{garcia2001}. 
Hence the question on whether or not electronic only effects are sufficient to explain LBMR and HBMR
remains.

Recently Garc\'{\i}a et al. \cite{garcia} have proposed that the presence of impurities, in particular oxygen, 
might be related to HBMR (exceeding 1,000 \%). This is an interesting hypothesis that concerns not
only experiments conducted in air but also those in ultra high vacuum conditions, since the samples usually
become contaminated after only a couple of hours.\cite{rego_rocha} Importantly at these extremely small
dimensions even a single impurity might have a large effect on the current flowing through the device. 
In particular, it is also well know that Ni is extremely reactive to oxygen and that bulk NiO is an insulator. 
Assuming that the insulating state of NiO persists in some form down to the atomic scale, then an oxygen 
contaminated Ni point contact can behave as an atomic scale tunnelling junction and perhaps sustain
LBMR or even HBMR. 

Large MR in ballistic tunnelling junctions are not uncommon in bulk junctions grown epitaxially \cite{parkin,yuasa}
and values in excess of 300\% have been demonstrated. Interestingly theoretical calculations have shown that 
the conductance is highly resonant in the two-dimensional Brillouin zone orthogonal to the transport 
direction.\cite{butler,ivanMgO} In particular, minority spins present high conductance in small regions away from 
the $\Gamma$ point and low conductance everywhere else. One may then speculate that such resonant
condition may be further exploited in quasi-1D system, where the transversal Brillouin zone collapses into a single point. 

Numerical simulations have an important role to play in addressing these issues. In particular it is necessary 
to determine with a high degree of accuracy the underlying electronic structure of the MPC and from that to 
calculate the electronic transport. This can be achieved by using density functional theory (DFT) \cite{hohen-kohn} 
in its Kohn-Sham form \cite{kohnsham} and the non-equilibrium Green function (NEGF) formalism.\cite{keldysh,datta}
However, for 3$d$ transition metals in presence of oxygen it is crucial to choose correctly the exchange and 
correlation (XC) potential, since this can significantly change the transport properties of the device.\cite{toher,toher1} 
Most notably, transition metal oxides are poorly described by the local spin density approximation (LSDA) and 
corrections are needed. A particularly useful one is offered by the LDA+$U$ scheme,\cite{anisimovldau1,anisimovldau2}
where the LSDA functional is replaced by an Hubbard-like energy for those atomic orbitals for which strong
correlation cannot be neglected (the 3$d$ shell in this case). 

This paper is organised as follows. In the next section we describe the method used to calculate the magnetoresistance 
as well as the two point contact arrangements investigated in our calculations. In section \ref{impfreeMPC} we present 
our results for the electronic transport properties of impurity-free MPCs using the LSDA and LDA+$U$ functionals.
In section \ref{OimpMPC} we discuss the role of oxygen impurities and finally we draw our conclusions.

\section{Method}\label{method}

Transport calculations are performed with {\it Smeagol} \cite{rochasmeagol,rocha1} our electronic transport code,
which uses a combination of DFT \cite{hohen-kohn,kohnsham} and NEGF method \cite{keldysh,datta} to accurate predict 
the $I$-$V$ characteristics of nano-scale devices. {\it Smeagol} is based on the DFT code SIESTA,\cite{siesta}
which provides the Kohn-Sham Hamiltonian over a basis set of numerical atomic orbitals.

\begin{figure}[ht]
\includegraphics[width=6cm,clip=true]{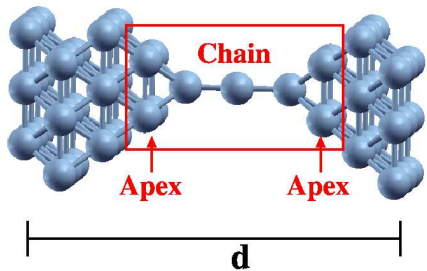}
\caption{(Color online) Ball and stick diagram of a nickel point contact formed by two pyramidal tips joined by a single Ni atom. 
The eleven atoms included in the red box correspond to the tip atoms (three atoms forming the atomic chain and 
four atoms on either side forming the apexes). $d$ is the distance between the two outermost planes.}\label{ballandstickPCatoms}
\end{figure}

\begin{figure}[h]
\center
\includegraphics[width=5.5cm,clip=true]{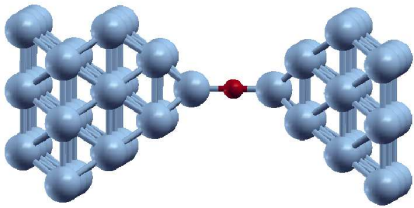}
\caption{(Color online) Ball and stick representation of the relaxed atomic arrangement of a nickel point contact with 
one oxygen atom brigding the gap between the two tips. Color code: grey (blue) Ni; black (red) O.}
\label{ballandstickoxygen}
\end{figure}

We ideally divide the device under investigation into three distinct regions, \cite{caroli} namely a left and a right 
semi-infinite electrode and a central scattering region where the potential drops.\cite{rochasmeagol} 
In the case of MPCs the scattering region consists of the atoms pictured in either Fig.\ref{ballandstickPCatoms} or 
Fig.\ref{ballandstickoxygen}. The main quantity in our transport calculation is the retarded Green function of the central scattering region evaluated at an energy $E$,
\begin{equation}
G^\sigma\left(E\right) = \lim_{\eta\rightarrow 0} \left[ \left(E+i\eta\right)S - H^\sigma - \Sigma^\sigma_\mathrm{L}\left(E\right) - \Sigma^\sigma_\mathrm{R}\left(E\right)\right]\:,
\end{equation}
where $H^\sigma$ is the Kohn-Sham Hamiltonian \cite{kohnsham} for spin $\sigma$ ($\sigma=\uparrow,\downarrow$), $S$ is the overlap matrix, $\Sigma^\sigma_\mathrm{L}$ ($\Sigma^\sigma_\mathrm{R}$) are the self-energys for the left and right lead respectively. The self energy contains information about the electronic structure of the semi-infinite electrode as well as the coupling to the scattering 
region.\cite{datta,rocha04,rochasmeagol,rocha1} Here we consider the two-spin fluid approximation,\cite{mott-sfluid} whereby the two spin components of the current add in parallel. In this particular case this approach is justified by calculations performed previously by Ying {\it et al.},\cite{maekawa} who successfully modeled a constricted junction using a mean-field Heisenberg model.

A self-consistent procedure based on the non-equilibrium Green function has been developed \cite{rocha04,rochasmeagol} for calculating both $G(E)$ and the charge density $\rho$. Once convergence has been achieved the energy- and bias-dependent transmission coefficients for each spin component can be calculated by using
\begin{equation}
T\left(E,V\right)^\sigma = \mathrm{Tr}\left[\Gamma_L G \Gamma_R G^\dagger \right]\:.
\end{equation}
with $\Gamma_\mathrm{L/R}\left(E\right)=i\left[\Sigma_\mathrm{L/R}-\Sigma_\mathrm{L/R}^\dagger\right]$. Finally, the spin 
polarised conductance is simply proportional to $T(E)$ evaluated at the Fermi level $E_\mathrm{F}$
\begin{equation}\label{conductance}
G = \frac{e^2}{h} \left[T^\uparrow\left(E_\mathrm{F}\right) + T^\downarrow\left(E_\mathrm{F}\right)\right]\:.
\end{equation}

We model the magnetoresistance in the MPC using the typical spin-valve scheme, i.e. we assume that in the absence of an external magnetic field, the magnetisation vectors of the two leads are opposite to each other in an antiparallel 
alignment (AA). Therefore, for zero field a sharp domain wall is formed inside the MPC.\cite{bruno} When a magnetic field is applied, the magnetic moments of the leads align in the field parallel to each other (PA) and the domain wall is eliminated from the junction. The low-bias ``optimistic" magnetoresistance ratio $R_\mathrm{GMR}$ can then be calculated by using the conductances of the AA ($G^\mathrm{AA}$) and the PA ($G^\mathrm{PA}$)
\begin{equation}\label{GMR}
R_\mathrm{GMR}=\frac{G^\mathrm{PA}-G^\mathrm{AA}}{G^\mathrm{AA}}\:.
\end{equation}

Our simulations are performed for atomic size point contacts, this arrangement 
 was initially proposed by Viret {\it et al.}.\cite{viret} The structure of a Ni break junction close to the rupture point is modelled by two Ni pyramids oriented along the [001] direction. These are formed from fcc bulk Ni, with a lattice constant equal to 3.46\AA\ and they sandwich one extra nickel atom in such a way that a three atom-long Ni chain bridges the two Ni surfaces (see figure \ref{ballandstickPCatoms}). The whole simulation cell consists of 56 nickel atoms and the basis set 
comprises of double-$\zeta$ (DZ) orbitals for the 4$s$, 4$p$ and 3$d$ orbitals, an additional polarisation orbital is included for the 4$s$.\cite{siesta} 
The real space grid is given by an equivalent plane wave cut-off of 500~Ry and we use 64 complex energy points for integrating the charge density.\cite{rochasmeagol} Finally we use periodic boundary conditions in the transverse direction and we sample over 8 $k$-points in the 2D Brillouin zone.

Before the transport calculations are carried out we perform conjugate gradient structure relaxation of 
the MPCs geometry. The two left-most and the two right-most atomic planes in the unit cell are kept fixed at the 
Ni fcc bulk positions, while the middle atoms are free to relax. Once the relaxation of a particular arrangement is 
completed we then increase the separation $d$ between the two outermost planes (see figure \ref{ballandstickPCatoms})
and another relaxation is performed. Thus we find the energy minimum as a function of $d$. The final distance 
between the plane of atoms forming the apex and the first atom in each side of the chain is 1.60 {\AA} and the inter-atomic distance within the chain itself is 2.24 {\AA}. Finally the distance between the plane of the apex and the first neighbouring plane of bulk atoms is 1.76 {\AA}.

We also consider the possibility of oxygen impurities in the MPC, by replacing the middle nickel atom 
of the single-atom chain by an oxygen atom (figure \ref{ballandstickoxygen}). In this case we perform structural 
relaxation to find the most energetically favourable arrangement of the device. For this calculation we use, in addition to 
the basis set of Ni, a double-$\zeta$ polarised basis for $s$ and $p$ orbitals of O. Atomic relaxation is performed from two
different initial geometries. The first has the oxygen atom placed along the axis of the two pyramids, while in the second 
this is displaced perpendicularly to the axis as a continuation of the fcc lattice. We find the same energy minimum for 
both the arrangements with the O atom in a straight configuration. 
In the transport calculations the relaxed coordinates are always considered. 

\section{Impurity-free MPC\lowercase{s}}\label{impfreeMPC}

\subsection{LSDA calculations}

\begin{figure}[h]
\center
\includegraphics[width=8cm,clip=true]{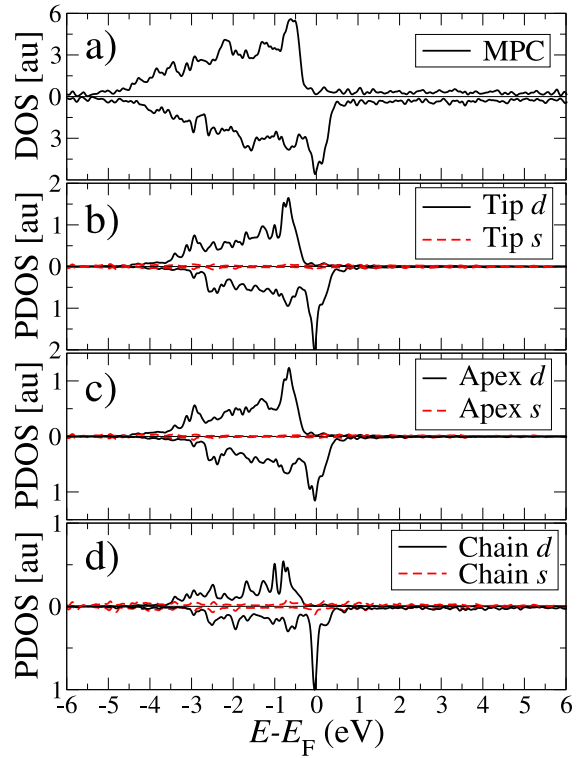}
\caption{(Color online) PDOS for impurity-free MPC: a) entire simulation cell, b) 11 atoms in the central region (apexes 
plus atomic chain), c) apex atoms, d) the 3 Ni atoms forming the monoatomic chain.}\label{NiPCPDOS}
\end{figure}

We start our analysis by looking at impurity-free MPCs. The projected density of states (PDOS)
can provide some initial insight into the character of the electronic states lying close to the Fermi level. 
In Fig.\ref{NiPCPDOS} we present the density of states projected onto the $s$ and $d$ orbitals of the entire MPC 
and on those of the atoms in the tip. 

\begin{figure}[h]
\center
\includegraphics[width=8cm,clip=true]{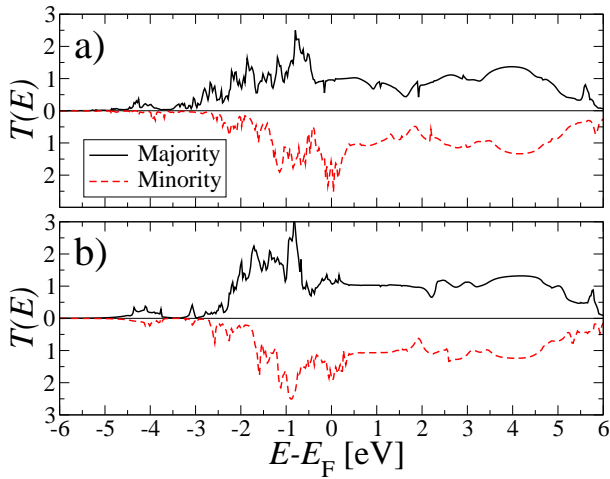}
\caption{(Color online) Transmission coefficients as a function of energy for Ni point contacts formed by a single atom chain: 
a) parallel and b) anti-parallel magnetic configuration.}\label{transmNiPC3atoms}
\end{figure}

First we note that the DOS of the entire MPC is similar to the one of the apexes and resembles that of bulk nickel.\cite{nickelmoment} In contrast the PDOS for the three-atom chain (Fig. \ref{NiPCPDOS}d) shows the effects of low-coordination, since it is much sharper than that of the rest of the MPC. We then expect peaks in the transmission 
coefficients. These are presented in Figure \ref{transmNiPC3atoms} for both the magnetic configurations of the MPC. 
In the PA configuration we observe minority-dominated transmission with $T^\downarrow(E_\mathrm{F})\approx$~2.5. 
The majority spins contribute with about $T^\uparrow(E_\mathrm{F})\approx 1$ to give a total conductance at zero bias around
 3.5~$e^2/h$. In the AA, albeit still dominant, the minority peak is largely suppressed and $T(E_\mathrm{F}$)
takes value similar to that of the majority spin in the PA. In fact, we observe a plateau at $T\left(E\right)\approx 1$ for 
$E > E_\mathrm{F}$ and both the spins. This originates from the presence of nearly un-polarised $s$ 
electrons whose transmission is rather insensitive to the magnetic configuration of the MPC.
The resulting MR ratio is approximately 20\%.

The presence of unpolarised $s$ electrons sets a serious limitation to the maximum MR achievable.
As we have just shown $s$ electrons contribute with about $2e^2/h$ to the conductance of both the AA and the PA. The 
remaining contribution originates from $d$ electrons. This is sensitive to the magnetic configuration, but
it cannot exceed $5e^2/h$ for each spin, since at most 5 $d$-bands can cross $E_\mathrm{F}$.
This gives us an upper bound for the MR of about 250\%. In practice however, $d$ electrons are strongly 
backscattered and they never contribute with more than 1.5 to the transmission coefficient. This indicates 
that huge MR is hardly possible in impurity-free MPC.

\subsection{LDA+$U$ Calculations}

Although useful and accurate in many situations, the LSDA severely underperforms for a variety of systems where 
strong electron correlations are important. Bulk Ni does not figure in this class of materials and in fact its electronic 
properties are extremely well described by the LSDA.\cite{nickelmoment} However the same cannot be said 
of Ni in atomic-size chains, where the localisation of the 3$d$ orbitals can be enhanced by the low coordination.
\begin{figure}[h]
\center
\includegraphics[width=5cm]{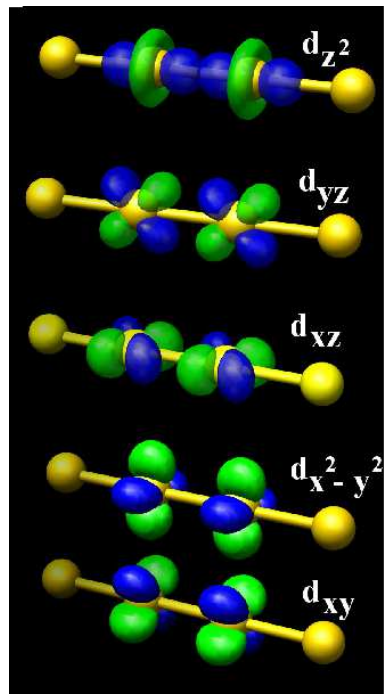}
\caption{(Color online) Schematic representation of the atomic orbitals responsible for the electronic structure of a Ni mono-atomic wire.
The $d_{x^2-y^2}$ and $d_{xy}$ orbitals are perpendicular to the axis of the wire and therefore weakly coupled.}\label{niwire}
\end{figure}

In figure \ref{niwire} we present a schematic diagram of the 3$d$ orbitals of a transition metal mono-atomic wire
aligned along the $z$ direction. While the $d_{z^2}$, $d_{yz}$ and $d_{xz}$ orbitals are parallel to the chain and
overlap considerably (note that in this arrangement symmetry forbids $s$-$d$ hybridisation for all the 3$d$ orbitals 
except for $d_{z^2}$). The $d_{xy}$ and $d_{x^2-y^2}$ are perpendicular to the chain and remain localised. 
The key point is that, in contrast to noble metals,\cite{rochasmeagol,rego_rocha} the localised $d$ orbitals in transition metals appear close to $E_\mathrm{F}$ and may affect drastically the low-bias electron transport. For this reason it is crucial to describe their position correctly, and therefore we need to consider XC potentials beyond LSDA.

Unfortunately, most of the corrections tend to deteriorate the LSDA electronic structure 
of bulk Ni. In the case of point contacts, the problem then becomes that of finding a good functional for 
the low connectivity apex region which at the same time does not alter the electronic structure of the planes at the 
edges of the cell. The LDA+$U$ method\cite{anisimovldau1,anisimovldau2} is particularly suited for this purpose. 
The approximation consists in replacing the LSDA XC energy with an on-site Hubbard-type
energy, which applies only to those atomic shells that need to be corrected (the $d$ shell in this case). The LDA+$U$
potential depends on two parameters, the Coulomb repulsion $U$ and the exchange $J$. 
These are associated to their corresponding atomic counterparts, although in a solid the atomic
values might be severely corrected by local screening. We can thus consider $U$ and $J$ non-vanishing only in the region
of the point contact where the Ni atoms have low coordination (the apexes and the tip).

In order to understand the effect of the LDA+$U$ on the electronic structure of the MPCs and to
extract realistic values for $U$ and $J$, we have first performed band-structure calculations for an infinite 
one-dimensional nickel chain (lattice spacing 2.24~{\AA}). For these calculations we have used the LSDA, LDA+$U$
with various choices of $U$ and $J$ and the atomic self-interaction corrected (ASIC) LSDA method.\cite{dasSIC}
ASIC is a fully {\it ab initio} scheme, which corrects for the atomic part of the LSDA self-interaction and it is proved
to perform extremely well for a broad range of materials including transition metal monoxides. 
\begin{figure}[h]
\includegraphics[width=9cm,clip=true]{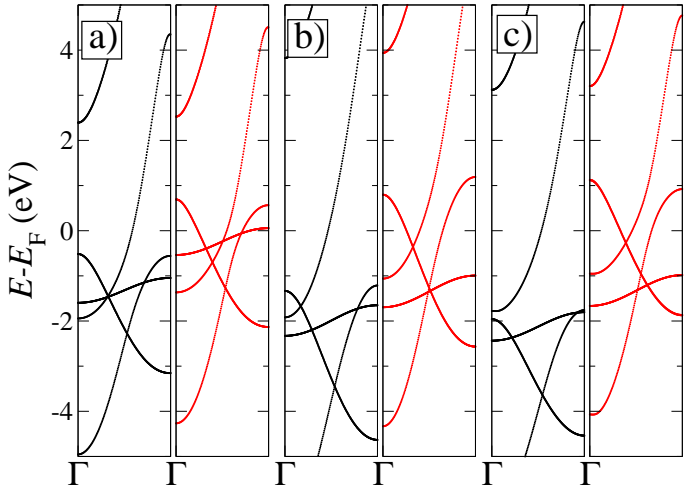}
\caption{(Color online) Band structures of a one-dimensional nickel chain (lattice spacing 2.24~{\AA}). a) LSDA, 
b) ASIC and c) LDA+$U$ with  $U$=6~eV and $J$=1~eV. The panels to the left (right) correspond to majority (minority) spins. }
\label{bandsNiwire}
\end{figure}
Recently it was also used for transport calculations \cite{toher,toher1} with considerable improvements over LSDA.
The bandstructure around the Fermi energy thus obtained are presented in figure \ref{bandsNiwire}. 

Generally speaking the 3$d$ manifold splits into three distinct bands, a broad hybrid $s$-$d_{z^2}$ band approximately 
8~eV wide and two doubly degenerates bands respectively of  $d_{yz}$, $d_{xz}$ and $d_{xy}$, $d_{x^2-y^2}$ symmetry. 
The first of these degenerate bands is about 2.5~eV wide (it is located at energies between -3~eV and -0.5~eV for the LSDA majority spin band), which is a direct consequence of the fact that the $d_{yz}$ and $d_{xz}$ orbitals are aligned along the chain.
In contrast the $d_{xy}$, $d_{x^2-y^2}$ band is only 0.5~eV wide (between -1.5~eV and -1~eV for the majority spin
of the LSDA bands). The spin-split is of about 1~eV similarly to bulk Ni. The main effect introduced by the LDA+$U$ and the ASIC is that the $d_{xy}$, $d_{x^2-y^2}$ band are shifted towards lower energies, thus moving them away from
$E_\mathrm{F}$. Taking the ASIC bands as reference we have fixed the Coulomb and exchange LDA+$U$
parameters respectively to $U$=6~eV and $J$=1~eV. Importantly we note that the $s$-$d_{z^2}$ band anti-crosses at around 
its band-centre and it is always present at the Fermi level regardless of the spin orientation and the XC functional used. 
One can thus expects a large portion of the conductance through the chain to be insensitive to the magnetic
state.

\begin{figure}[h]
\includegraphics[width=8cm,clip=true]{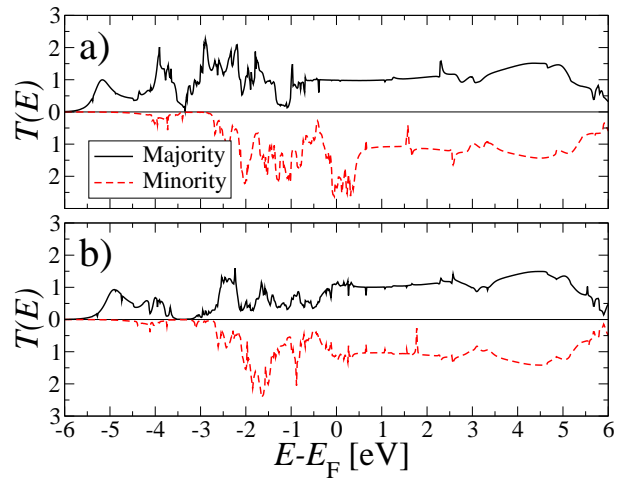}
\caption{(Color online) Transmission coefficients as a function of energy using LDA+$U$ for Ni MPC. a) Parallel and b) anti-parallel 
alignments of the magnetic moments of the electrodes. The values of $U$ and $J$ are 6~eV and 1~eV respectively.}\label{transmNiPC3atomsU}
\end{figure}

We then move to the calculation of the transport properties of the MPC of Fig. \ref{ballandstickPCatoms}.  Here
we apply the LDA+$U$ corrections only to the three atoms forming the mono-atomic chain. Test calculations where 
the corrections are extended to the apexes yield essentially the same results. The zero-bias transmission coefficients
for the PA and AA and for the different spins are presented in figure \ref{transmNiPC3atomsU}, little difference is observed when the transmission coefficients are calculated with LSDA (see Fig.\ref{transmNiPC3atoms}). In the PA case the transmission for the majority spins are close to unity for a wide range of energies, indicating less hybridization between $s$ and $d$ orbitals, while the minority spin still has a peak of $T\approx 2.5$ at $E_\mathrm{F}$. In general the main difference with the LSDA results is that now the transmission for energy below $E_\mathrm{F}$ is reduced and the peaks are broadened. This is the result of the downshift of the $d$ orbitals due to the on-site $U$ correction.

Also the transmission for the AA is  similar to that calculated with LSDA, except for a general reduction of $T$. This is now rather close to unity at the Fermi energy for both the spin configurations (note that since the domain wall is located between the second and third atom in the chain, the two spin directions are not degenerate for the AA).

The results of figure \ref{transmNiPC3atomsU} can be understood by noting that the MPC investigated is essentially
formed by a linear mono-atomic Ni chain sandwiched between two Ni surfaces. The transmission coefficient thus
reflects closely the bandstructure of the chain itself. Consider first the majority spins for the PA. The large plateau in $T^\uparrow\left(E\right)$
for $E\ge E_\mathrm{F}$ is associated with the broad $s$-$d_{z^2}$ hybrid band. This is the only one present in the
bandstructure of the Ni chain in that energy range. Below $E_\mathrm{F}$ the contribution of the remaining $d$ orbitals 
becomes important and $T$ exceeds unity. Such a contribution is present at $E_\mathrm{F}$ for the minority spins, for 
which 4 bands cross the Fermi level. One thus expects the conductance of the minority spins to reach up to $4e^2/h$. 
However scattering with the pyramid-shape electrodes results in a value of only about $2.5e^2/h$ in the 
actual MPC.

Similarly the transmission for the antiparallel configuration can also be related to the bands of Fig. \ref{bandsNiwire}.
In this case the MPC can be thought of as two Ni chains with opposite magnetization direction, coupled to each other 
through an atomically sharp domain wall. Therefore electrons propagating in the majority (minority) band across
the first half of the chain, must propagate as minority (majority) in the other half.\cite{rocha04} Consequently the transmission is large at those energies where bands with the same orbital character appear for both spin directions. The only 
band that satisfies this criterion for energies around $E_\mathrm{F}$ is the hybrid $s$-$d_{z^2}$, which explains why 
$T(E_\mathrm{F})\sim 1$ for the AA. In contrast the remaining ``pure'' $d$ bands are narrow and exchange split 
and contribute little to the total transmission. 

The GMR in this case is enhanced with respect to that calculated with LSDA and can reach up 60~\%. This is 
in agreement with calculations performed with different {\it ab initio} methods~\cite{mertig1,palaciosNi} and 
experiments for atomically thin junctions.\cite{viret} However, this value is still far from those observed in 
some of the experiments where the HBMR is measured.\cite{garcia2001,zchopra}
The broad $s$-$d_{z^2}$ hybrid band is both highly conductive and crosses the Fermi energy for the two spin sub-bands. This makes it difficult to compare our {\it ab initio} results with that class of experiments.
Moreover it is also important to note that the dominant $s$ orbital character of the $s$-$d_{z^2}$ band 
makes such conductance channel rather robust against conformal changes of the point contact. In fact
calculations with different MPC geometries give rather similar results. Thus, one must conclude that impurity 
free MPCs can not produce HBMR. 

Next we consider the case of MPCs contaminated with strongly electronegative impurities 
such as oxygen. Our expectation is then to remove the contribution to the conductance originating 
from the $s$ orbitals.

\section{Oxygenated N\lowercase{i} MPC\lowercase{s}}\label{OimpMPC}

\subsection{LSDA Calculations}

Following the suggestions of Garcia et al. \cite{garcia} we consider the effects of oxygen contamination 
in MPCs. We start by calculating the LSDA bandstructure of a mono-atomic NiO linear chain. In doing this we
follow the same idea of the previous section, i.e. that the transport properties of the MPC can be understood
from the electronic structure of the one dimensional chain associated to the three atoms forming
the narrower part of the constriction. 
\begin{figure}[h]
\center
\includegraphics[width=7cm,clip=true]{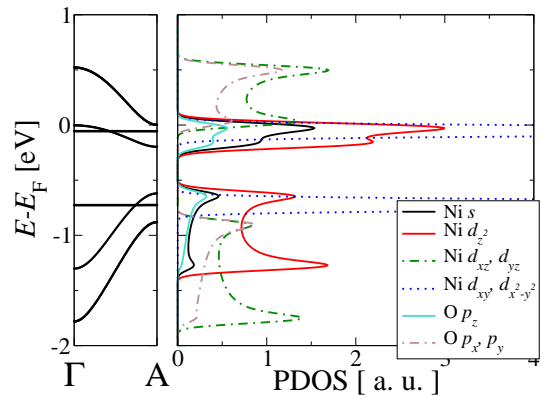}
\caption{(Color online) Band structures (left panel) and projected density of states (right panel) for an infinite 
one-dimensional NiO wire in the antiferromagnetic configuration calculated with the LSDA. In this case the
two spin sub-bands are degenerate.}\label{NiO1DbandsLDAAA}
\end{figure} 

\begin{figure}[h]
\includegraphics[width=8cm,clip=true]{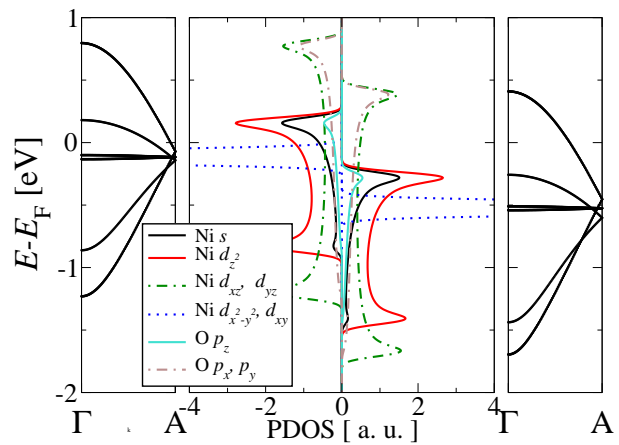}
\caption{(Color online) Band structures and projected density of states for an infinite one-dimensional NiO wire in the ferromagnetic 
configuration calculated with the LSDA. The left (right) panels are for the minority (majority) spins.}\label{NiO1DbandsLDAPP}
\end{figure}

In this case the chain associated to the MPC of figure \ref{ballandstickoxygen} is a NiO chain with a bond length of
1.8 {\AA}. The bandstructure and the corresponding PDOS for the antiferromagnetic and ferromagnetic ground state 
are presented respectively in figures \ref{NiO1DbandsLDAAA} and \ref{NiO1DbandsLDAPP}.
The LSDA total energies of these two configurations are identical, which means that the chain is actually paramagnetic.
Metallicity is found regardless of the magnetic state and it is due to a hybrid $s$-$d_{z^2}$ band. This band receives contributions 
from the O $p_z$ orbitals and has a spin-split of about 0.5~eV in the 
ferromagnetic configuration with the majority component completely filled. Thus, if one assumes that the 
transport through the chain is completely dominated by such a highly conducting band, then the ferromagnetic
one-dimensional NiO will effectively behave as a half-metal with a potentially large GMR. 
However in the antiferromagnetic configuration such band splits into two narrow bands respectively 
at the Fermi level and at about 0.6~eV below $E_\mathrm{F}$, precluding the possibility of a strong
suppression of the current for the AA.

\begin{figure}[h]
\center
\includegraphics[width=8cm,clip=true]{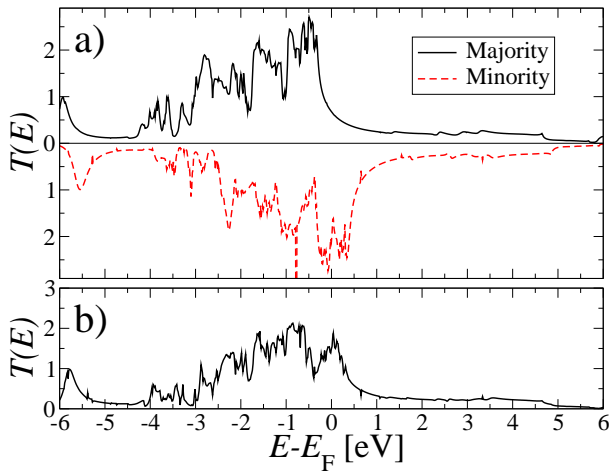}
\caption{(Color online) Zero-bias transmission coefficients of oxygen contaminated Ni point contacts calculated with the LSDA:  a) parallel and b) anti-parallel alignments. In the PA the solid (dashed) line represents the majority (minority) spins, while the two spins are degenerate in the AA. }\label{transmNiPC_O}
\end{figure}

With these results in hand we move on to calculating $T(E)$ for the structure shown in figure 
\ref{ballandstickoxygen}, these results are shown in figure \ref{transmNiPC_O}. The transmission coefficients 
at the $E_\mathrm{F}$ for the PA configuration are large for the minority and small for the majority spins 
(respectively 0.5 and 2.5), this is consistent with our discussion based on the NiO chain. 
However $T(E_\mathrm{F})$ is rather large for both spins in the AA because of the contribution from the 
Ni $s$-$d_{z^2}$-O $p_z$ band. This gives an overall conductance larger than that of the PA and a
tiny negative GMR of approximately -6 \%. It therefore appears that oxygen impurities are unable to explain HBMR.

This negative result however should be taken with some care. As mentioned in the introduction LSDA poorly reproduces the electronic structure of standard Mott-Hubbard insulators such as 
NiO. One may argue that the same problem appears in these low dimensional oxides and therefore our 
results need to be tested against the LDA+$U$ scheme. 

\subsection{LDA+$U$ calculations}

As in the previous sections we begin with the LDA+$U$ calculation of an infinite NiO chain. Here we 
consider the same values of $U$ and $J$ used for the case of the Ni-only chains. The configuration 
presenting the lowest energy turns out to be non-magnetic,\footnote{The non-magnetic solution 
is the most favourable for a range of values of $U$ between 2 eV to 8 eV ($J$ is kept constat at 1~eV).} 
although both ferromagnetic and antiferromagnetic solutions can be stabilized. Interestingly
several antiferromagnetic states, both conducting and insulating, have been obtained with the convergence 
being extremely sensitive to the initial orbital occupation. 

\begin{figure}[h]
\center
\includegraphics[width=8cm]{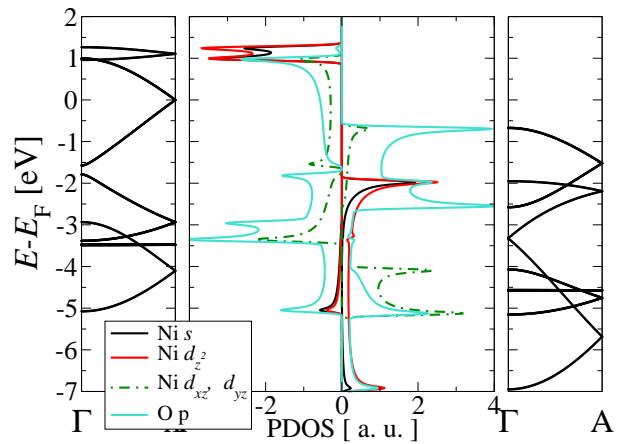}
\caption{(Color online) Band structures and projected density of states for an infinite one-dimensional NiO wire in the ferromagnetic 
configuration calculated with the LDA+$U$. The left (right) panels are for the minority (majority) spins. 
The values of $U$ and $J$ are set to 6~eV and 1~eV respectively.}\label{NiO1DbandsLDAUFM}
\end{figure}

In figure \ref{NiO1DbandsLDAUFM} we present the bandstructure for the ferromagnetic configuration. 
These appear significantly different from their LSDA counterpart, since NiO is now a half-metal. Moreover the $s$ electrons
are no longer present at the Fermi level, which in contrast is characterized by orbitals orthogonal to the chain axis,
namely the oxygen $p_x$ and $p_y$ and the Ni $d_{xz}$ and $d_{yz}$. 

\begin{figure}[h]
\includegraphics[width=7cm,clip=true]{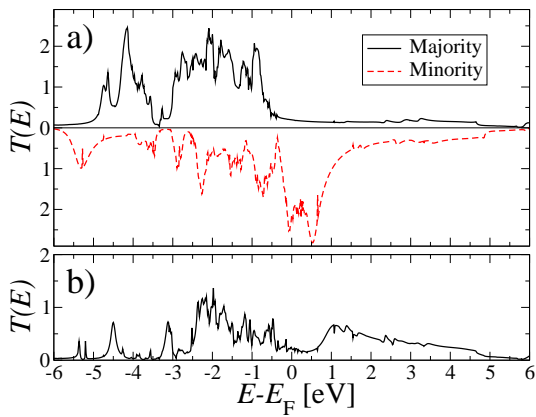}
\caption{(Color online) Zero-bias transmission coefficients of oxygen contaminated Ni point contacts calculated with the LDA+$U$:  a) parallel and b) anti-parallel alignments. In the PA the solid (dashed) line represents the majority (minority) spins, while the two spins are degenerate in the AA. }\label{transmNiPC_O_LDAU}
\end{figure}

We then proceed with calculating the transport through the point contact. When the structure of figure \ref{ballandstickoxygen}
is considered the non-magnetic solution is no longer found. The transmission 
coefficients calculated using LDA+$U$ are shown in figure \ref{transmNiPC_O_LDAU}. The PA configuration is qualitatively 
similar to the LSDA result (see figure \ref{transmNiPC_O}a) however the transmission for minority spins is slightly higher while that for majority is lower ($T\sim 0.2$). The majority $d$ states are lower than in the LSDA case and although the transmission coefficients at $E_\mathrm{F}$ are non-zero it does resemble the 
half-metal behaviour observed in the infinite chain. The residual conductance at $E_\mathrm{F}$ is then attributed to
direct tunnelling at the apexes, which is substantial for these small separations. 

For the AA configuration the LDA+$U$ results are in stark contrast to the LSDA ones and in particular there is a drastic
reduction of the transmission at $E_\mathrm{F}$. This is somehow expected since the contribution of the $s$ electrons
to the transmission has been severely reduced. The resulting magnetoresistance reaches a much higher value of 450~\%
in good agreement with results obtained with by using hybrid XC functionals.\cite{palaciosNiO}

\section{Conclusion}

In summary, we have investigated the possibility of LBMR and HBMR in Ni point contacts using a combination of
DFT and NEGF method. In particular we have explored the dependence of the MR on the XC functionals used
and the effects of oxygen contamination. For defect-free point contacts both LSDA and and LDA+$U$ calculations
agree in setting an upper bound to the MR in the range of 50~\%. 
This rather small value is essentially due to the high transmission associated to an unpolarized $s$-$d_{z^2}$ 
band, which is always present at the Fermi level. This is rather insensitive to conformal changes of the MPC and
contributes $2~e^2/h$ to the conductance regardless of the magnetic state of the device. 

In contrast, oxygenation can drastically change the picture. In particular O $p$ states can hybridize with the $s$-$d_{z^2}$
band shifting it away from $E_\mathrm{F}$. However, in this case the results are  strongly dependent on the exchange and
correlation functional, with LSDA giving a small negative MR while LDA+$U$ gives a positive MR of about 450~\%. 
Since the well-known problem of LSDA of dealing with transition metal monoxides, we believe that the LDA+$U$ results 
are the most relevant. Importantly even in this case the MR is still much smaller than those measured in
experiments showing HBMR. The main point is that, although strongly suppressed, there is still substantial current
for the AA, attributed to direct tunnelling of the apexes. Certainly larger oxygenation, or the formation of long NiO 
chains may suppress this leakage current, however we believe that highly oxygenated configurations are rather unlikely. 
This leads us to conclude that HBMR may have additional origins besides electronic considerations.

\section{Acknowledgments}

We would like to thank Chaitanya Das Pemmaraju for stimulating discussions. This work is supported by Science 
Foundation of Ireland under the grants SFI02/IN1/I175 and SFI05/RFP/PHY0062. Computational resources have 
been provided by the HEA IITAC project managed by the Trinity Centre for High Performance Computing and by ICHEC.

\end{document}